# Commutativity of Systems with their Feedback Conjugates

Mehmet Emir KOKSAL

*Department of Mathematics, Ondokuz Mayis University, 55139 Atakum, Samsun, Turkey*

*emir_koksal@hotmail.com*

**Abstract:** After introducing commutativity concept and summarizing the relevant literature, this work is focused on the commutativity of feedback conjugates. It is already known that a linear time-varying differential system describing a single input-single output dynamical system is always commutative with its constant gain feedback pairs. In this article, it is proven that among the time-varying feedback conjugates of a linear time-varying system, constant feedback conjugates are the only commutative feedback pairs and any of the time-varying feedback conjugates cannot constitutes a commutative pair of a linear time-varying system.

1. Introduction

Consider a linear time-varying $N^{th}$ order $(N \geq 1)$ differential system $A$ described by

$$\sum_{n=0}^{N} a_n(t) \frac{d^n}{dt^n} y_A(t) = x_A(t) \tag{1a}$$

where $x_A(t)$ is the input, $y_A(t)$ is the output of the system; $a_n(t)$ are time-varying coefficients with $a_N(t) \neq 0$ for $\forall t \geq t_0 \in R$; $y_A^{(n)}(t_0) = y_{An} \in R$ are the constant initial conditions of the output and it's derivatives up to $(N-1)^{th}$ order at some initial time $t_0 \in R$. Further, $a_n(t)$ and $x_A(t)$ are piecewise continuous functions from $[t_0, \infty) \to R$; this function space is denoted by $P[t_0, \infty)$. It is well known that this system has a unique solution for the output $y_A(t)$ and it describes a dynamical system [1].

Next consider an alternate system B which is similar to Eq. (1a) with order $M$, $1 \leq M \leq N$;

$$\sum_{m=0}^{M} b_m(t) \frac{d^m}{dt^m} y_B(t) = x_B(t) \tag{1b}$$

where different parameters are defined as in (1a).

When the systems A and B are connected one after the other (cascade or chain connection [2]) to form a larger system AB or BA with input $x$ and output $y$ ($x = x_A, y = y_B, y_A = x_B$ for $AB$; $x = x_B, y = y_A, y_B = x_A$ for $BA$), then if both of the connections AB and BA have the same input-output pairs (x, y) for all inputs $x \in P[t_0, \infty)$, we say that A and B are commutative and $(A, B)$ constitutes a commutative pair.

The concept of commutativity was first introduced by J. E Marshall in 1977 [3]. Marshal investigated the commutativity conditions of the systems of order less or equal to one, further he proved that for commutativity, either both systems are time-invariant or both systems are time-varying. This general statement excludes zero-order systems which are always commutative pairs among themselves whether they have time-invariant or time-varying gains; in the case one of the systems is a zero order system with constant scalar gain, such a system is commutative with all time-varying systems of any order.

Later Marshall's work of commutativity, considerable and continuing research has been done on commutativity. In [4-6], the necessary and sufficient conditions for commutativity of second order systems are presented. Koksal has presented the general commutativity conditions for time-varying systems of any order and reformulated the previous results obtained for $2^{nd}$ order systems in the format of general conditions [7]. In this work, the general conditions are used to show that any system with constant forward and feedback gains is commutative with the system itself, which is an important fact for the feedback control theory.

In 1985, Koksal prepared a technical report which is a survey on commutativity [8]. This report being summarizing all the previous results has the following original contributions: i) An iterative formula is derived for the entries of the coefficient matrix expressing the first set of commutativity conditions. ii) An explicit formula is given for the entries of the coefficient matrix expressing the second set of commutativity conditions, and thus the theorem stating these conditions is formally proved. iii) Explicit commutativity results for $4^{th}$ order systems are obtained. iv) Finally, commutativity of Euler's system is proved.

The content of the published but undistributed work [8] can be found in journal paper [9] which presents an exhaustive study on the commutativity of time-varying systems. This paper is the first tutorial paper that has appeared in the literature.

More than one decade no publications had been appeared in the literature until the work in 2011 [11]. This reference is the second basic journal publication after the first appeared in 1988 [9]. It covers a precise summary of the previous results some of which had not been announced widely being in an unpublished report or in a national/international conference proceeding [10]. Commutativity in case of nonzero initial conditions, commutativity of Euler's systems, new results about the effects of commutativity on sensitivity, the reduction of disturbance by change of connection order in a chain structure of subsystems and the most important the explicit commutativity conditions for $5^{th}$ order systems which have never been published before anywhere are among the subtitles of this reference.

About the commutativity of continuous time linear time-varying systems, [12] has been the last paper appearing in the literature; it deals with finding all second order commutative pairs of a first order linear time-varying system.

Commutativity is introduced and several publications have appeared for discrete-time systems as well [13, 14]. However, the commutativity of discrete-time systems is out of the scope of this contribution.

## 2. Feedback Connection

Consider a dynamical system $A$ with input $x_A(t)$ and output $y_A(t)$ simply shown in Fig. (1a). Its feedback conjugate $B$ involving forward path gain $\alpha(t)$ and feedback path gain $\beta(t)$ is shown in Fig. (1b). Although feedback control theory involves many alternatives of feedback structures and multiple feedback loops [15-17], the scope in this paper will be confined to the basic feedback conjugate shown in this figure. It is assumed that $A$ and $B$ are represented by the differential systems (1a) and (1b), respectively. Due to the connection shown in Fig. (1b), the constraint equations

a)

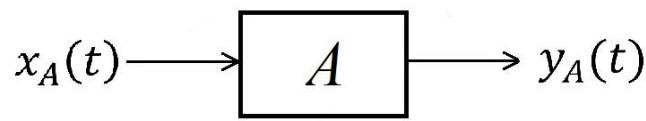

b)

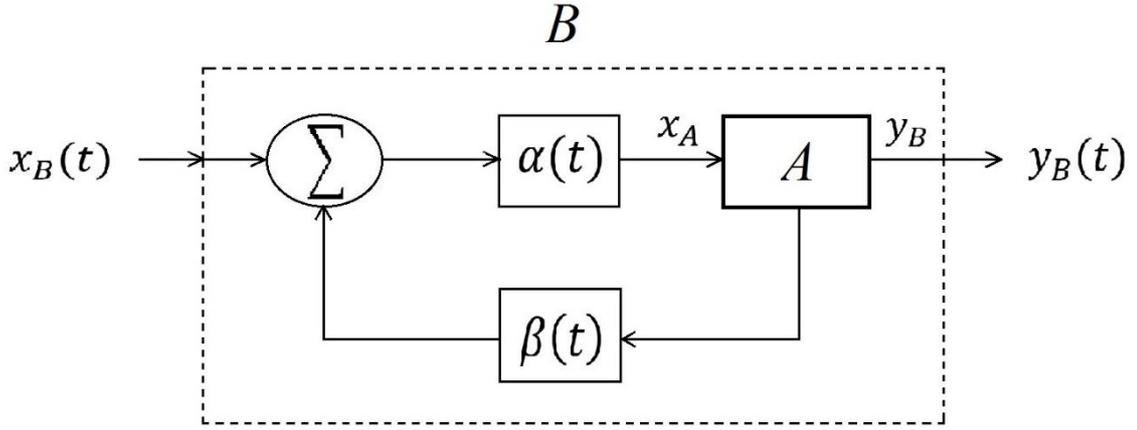

**Figure 1:** a) Dynamical system $A$ and b) its feedback conjugate $B$.

$$y_A(t) = y_B(t), \qquad (2a)$$

$$x_A(t) = \alpha(t)[x_B(t) - \beta(t)y_A(t)] = \alpha(t)x_B(t) - \alpha(t)\beta(t)y_B(t) \qquad (2b)$$

reduce the representation of Eq. (1a) into the following form

$$\sum_{n=1}^{N} \frac{a_n(t)}{\alpha(t)} \frac{d^n}{dt^n} y_B(t) + \left[\frac{a_0(t)}{\alpha(t)} + \beta(t)\right] y_B(t) = x_B(t). \qquad (3)$$

This equation is in the form of Eq. (1b) with $M = N$ and

$$b_n(t) = \frac{a_n(t)}{\alpha(t)} \quad for\ n = N, N-1, \cdots, 1 \qquad (4a)$$

$$b_0(t) = \frac{a_0(t)}{\alpha(t)} + \beta(t) \quad for\ n = 0. \qquad (4b)$$

Since any constant gain scalar (order 0) system is commutative with any linear time-varying system and two variable gain scalar systems are commutative among themselves, the scope of this presentation is devoted to commutativity of two time-varying systems at least one of them has order equal or greater than 1. Note that the subsystem A and its feedback conjugates of type $B$ shown in Fig. (1b) are of the same order $N$, and within the defined scope $N \geq 1$; further, it is assumed that $\alpha(t) \neq 0$ identically for it leads to a trivial system $B$ having zero output, and $\alpha(t) \neq 0$ for $\forall t \geq t_0$ to prevent the ambiguity to be appeared in Eq. (3) and Eq. (4).

## 3. Commutativity of Variable Feedback Conjugates

It has been proven that any two identical linear time-varying systems with arbitrary time-invariant forward and feedback path gains are commutative [9, 11]. This the statement that any time-varying system is commutative with its feedback conjugates having constant forward and feedback gains since any feedback system reduces to its original with unit forward and zero feedback gains.

For a scalar system $A$, that is for systems of order 0, Eq. (1a) indicates a variable gain scalar system and its feedback conjugate B represented by Eq. (3) is also a variable gain scalar system. Since scalar systems are always commutative, any zero order system is always commutative with its time-varying feedback conjugates. Further Eq. (4b) implies that a time-varying scalar system ($a_0(t) \neq 0$) may have a time-invariant commutative feedback pair (let $\alpha(t) = -a_0(t)$ and $\beta(t) = constant$ for example). The case of scalar (zero order) systems are excluded from Marshall's Theorem [3] and Corollary 1 of [9] stating "For commutativity of two linear systems, it is required that either both systems are time-invariant or both system are time-varying"; more strikingly "A time-varying system can never commute with a time-invariant system". Therefore the scope of the sequel is kept in mind for systems of order 1 or greater.

We now return to the inverse problem that; is the time invariance of the forward and feedback gains necessary for a linear time-varying system for being commutative with its feedback conjugates?

For simplicity consider the problem for a first order system $A$ of which feedback conjugates are described by a first order differential system in the form of Eq. (3) where $N = 1$ and with coefficients in Eq. (4). When $b_1$ and $b_0$ in Eq. (4) are replaced in the commutativity condition (9) in [11], we obtain

$$\begin{bmatrix} \dfrac{a_1(t)}{\alpha(t)} \\ \dfrac{a_0(t)}{\alpha(t)} + \beta(t) \end{bmatrix} = \begin{bmatrix} a_1(t) & 0 \\ a_0(t) & 1 \end{bmatrix} \begin{bmatrix} c_1 \\ c_0 \end{bmatrix}. \tag{5}$$

Since $a_1(t) \neq 0$ and $\alpha(t) \neq 0$ identically, the first equation implies that $\alpha(t) = 1/c_1$ which is a constant. Further, the second equation implies $\beta(t) = c_0$ which is also constant. Hence, for all the commutative feedback conjugates of $A$, the forward path and the feedback path gains are constant.

For a second order system, that is for $N = 2$, when the coefficients of the feedback conjugates in Eq. (4) are substituted in the commutativity condition expressed by (10a) in [9], the following results

$$\begin{bmatrix} \dfrac{a_2(t)}{\alpha(t)} \\ \dfrac{a_1(t)}{\alpha(t)} \\ \dfrac{a_0(t)}{\alpha(t)} + \beta(t) \end{bmatrix} = \begin{bmatrix} a_2(t) & 0 & 0 \\ a_1(t) & a_2^{0.5}(t) & 0 \\ a_0(t) & \dfrac{a_2^{-0.5}(t)[2a_1(t) - \dot{a}_2(t)]}{4} & 1 \end{bmatrix} \begin{bmatrix} c_2 \\ c_1 \\ c_0 \end{bmatrix}. \tag{6}$$

Since $a_2(t) \neq 0$ and $\alpha(t) \neq 0$ identically, the first row implies $\alpha(t) = 1/c_2$. Then, the second row is satisfied by choosing $c_1 = 0$. Finally the last row yields $\beta(t) = c_0$. Hence, it is obvious that the forward and feedback path gains cannot vary with time and they are necessarily constant.

The above observations for first and second order systems are valid for higher order systems as well. We express this fact by a theorem and prove it.

**Theorem 1:** *Consider a non-scalar linear time-varying system described by an $N^{th}$ order differential system in the form (1). Among all of its feedback conjugates with time-varying forward and feedback gains depicted in Fig. (1b), only the ones having constant forward* *and feedback gains are commutative with the original system.*

**Proof:** The necessary conditions for commutativity of two $N^{th}$ order systems are expressed in Eq. (5c) in [9]. When the coefficients of the feedback system as expressed in Eq. (4) are substituted in this equation we obtain

$$\begin{bmatrix} \frac{a_N(t)}{\alpha(t)} \\ \frac{a_{N-1}(t)}{\alpha(t)} \\ \frac{a_{N-2}(t)}{\alpha(t)} \\ \vdots \\ \frac{a_1(t)}{\alpha(t)} \\ \frac{a_0(t)}{\alpha(t)} + \beta(t) \end{bmatrix}$$

$$= \begin{bmatrix} a_N(t) & 0 & 0 & 0 & \cdots & 0 & 0 \\ a_{N-1}(t) & a_N^{\frac{N-1}{N}}(t) & 0 & 0 & \cdots & 0 & 0 \\ a_{N-2}(t) & f_{N-2,N-1}(t) & a_N^{\frac{N-2}{N}}(t) & 0 & \cdots & 0 & 0 \\ \vdots & \vdots & \vdots & \vdots & \ddots & \vdots & \vdots \\ a_1(t) & f_{1,N-1}(t) & f_{1,N-2}(t) & f_{1,N-3}(t) & \cdots & a_N^{\frac{1}{N}}(t) & 0 \\ a_0(t) & f_{0,N-1}(t) & f_{0,N-2}(t) & f_{0,N-3}(t) & \cdots & f_{0,1}(t) & 1 \end{bmatrix} \begin{bmatrix} c_N \\ c_{N-1} \\ c_{N-2} \\ \vdots \\ c_1 \\ c_0 \end{bmatrix} \quad (7)$$

Since $a_N(t) \neq 0$, the first equation implies $\alpha(t) = 1/c_N$ which is a constant, with this the second equation implies $c_{N-1} = 0$. Proceeding sequentially until the last equation we obtain $c_{N-2} = 0, c_{N-3} = 0, \cdots, c_1 = 0$. Finally, the last equation yields $\beta(t) = c_0$ which is also a constant. Hence, the proof is completed.

Next, we consider the inverse of Corollary 3 in [9] which states that any two identical $N^{th}$ order time-varying systems with arbitrary time-invariant forward and feedback path gains are commutative. It is true that the necessity of being constant for path gains is not required for commutativity of time-varying feedback conjugates. We state this result explicitly by the following theorem.

**Theorem 2:** *For commutativity of any two time-varying feedback conjugates of an $N^{th}$ order time-varying linear system, it is not necessary that their forward and feedback path gains are constant; they should satisfy*

$$\alpha_2(t) = p\alpha_1(t), \qquad (8a)$$

$$\beta_2(t) = p\beta_1(t) + q, \qquad (8b)$$

*where $\alpha_1(t)$, $\alpha_2(t)$ are forward path gains; $\beta_1(t)$, $\beta_2(t)$ are feedback path gains of the two time-varying feedback conjugates; and $p \neq 0$ and $q$ are some constants.*

**Proof:** Let $B_1$ and $B_2$ be two time-varying feedback conjugates of A in Fig. (1) have forward gains $\alpha_1(t)$ and $\alpha_2(t)$, feedback gains $\beta_1(t)$, and $\beta_2(t)$, respectively. Rewriting Eq. (3) for $B_1$ and $B_2$, we have

$$\sum_{n=1}^{N} \frac{a_n(t)}{\alpha_1(t)} \frac{d^n}{dt^n} y_{B_1}(t) + \left[\frac{a_0(t)}{\alpha_1(t)} + \beta_1(t)\right] y_{B_1}(t) = x_{B_1}, \qquad (9a)$$

$$\sum_{n=1}^{N} \frac{a_n(t)}{\alpha_2(t)} \frac{d^n}{dt^n} y_{B_2}(t) + \left[\frac{a_0(t)}{\alpha_2(t)} + \beta_2(t)\right] y_{B_2}(t) = x_{B_2}. \qquad (9b)$$

Using the necessary condition expressed by Eq. (5c) in [9] by considering the system A and B as $B_1$ and $B_2$, respectively, for commutativity of $B_1$ and $B_2$ we obtain

$$= \begin{bmatrix} \dfrac{a_N(t)}{\alpha_2(t)} \\ \dfrac{a_{N-1}(t)}{\alpha_2(t)} \\ \dfrac{a_{N-2}(t)}{\alpha(t)} \\ \vdots \\ \dfrac{a_1(t)}{\alpha_2(t)} \\ \dfrac{a_0(t)}{\alpha_2(t)} + \beta_2(t) \end{bmatrix}$$

$$= \begin{bmatrix} a_N(t) & 0 & 0 & 0 & \cdots & 0 & 0 \\ \dfrac{a_{N-1}(t)}{\alpha_1(t)} & \left[\dfrac{a_N(t)}{\alpha_1(t)}\right]^{\frac{N-1}{N}} & 0 & 0 & \cdots & 0 & 0 \\ \dfrac{a_{N-2}(t)}{\alpha_1(t)} & f_{N-2,N-1}(t) & \left[\dfrac{a_N(t)}{\alpha_1(t)}\right]^{\frac{N-2}{N}} & 0 & \cdots & 0 & 0 \\ \vdots & \vdots & \vdots & \vdots & \ddots & \vdots & \vdots \\ \dfrac{a_1(t)}{\alpha_1(t)} & f_{1,N-1}(t) & f_{1,N-2}(t) & f_{1,N-3}(t) & \cdots & \left[\dfrac{a_N(t)}{\alpha_1(t)}\right]^{\frac{1}{N}} & 0 \\ \dfrac{a_0(t)}{\alpha_1(t)} + \beta_1(t) & f_{0,N-1}(t) & f_{0,N-2}(t) & f_{0,N-3}(t) & \cdots & f_{0,1}(t) & 1 \end{bmatrix} \begin{bmatrix} c_N \\ c_{N-1} \\ c_{N-2} \\ \vdots \\ c_1 \\ c_0 \end{bmatrix} \quad (10)$$

The first row of equation in (10) implies

$$\frac{1}{\alpha_2(t)} = \frac{c_N}{\alpha_1(t)}. \tag{11a}$$

The second equation yields $c_{N-1} = 0$. Proceeding downward through the rows and using the results previously obtained, it is found that $c_{N-2} = 0, c_{N-3} = 0, \cdots, c_1 = 0$. Finally substituting these values in the last row of equation we obtain

$$\beta_2(t) = \beta_1(t)c_N + c_0. \tag{11b}$$

Finally comparing Eq. (8) and Eq. (11) it is seen that Theorem 2 is proved and $p = 1/c_N$ and $q = c_0$ are the proper constants.

## 4. Conclusions

A short but comprehensive literature survey is given on commutativity. Some details about the commutativity of non-scalar linear time-varying differential systems are considered. Two theorems which complete the voids in the literature on the subject are stated and proved.

The first theorem shows that a non-scalar linear time-varying system and any of its feedback conjugates with time-varying forward and feedback path gains are commutative if and only if forward and feedback path gains are time invariant.

The second theorem reveals that commutativity of feedback conjugates of a non-scalar linear time-varying system may or may not require constant forward and feedback gains, but if they are time-varying the path gains cannot change independently so that they are interrelated as in Eq. (8).

The theorems stated exclude constant or time-varying scalar systems which are always commutative. However the case of non-zero initial conditions is not included in this paper and this stems for another research subject.

**Acknowledgments:** This study is supported by the Scientific and Technological Research Council of Turkey (TUBITAK) under the project no. 115E952.